\documentclass{pasj00}

\begin{document}
\SetRunningHead{TOTANI}{The Galactic Center Black Hole}

\title{A RIAF Interpretation for 
the Past Higher Activity of the Galactic Center Black Hole 
and the 511 keV Annihilation Emission }

\author{Tomonori \textsc{TOTANI}}%
\affil{Department of Astronomy, Kyoto University, Sakyo-ku, 
  Kyoto 606-8502}
\email{totani@kusastro.kyoto-u.ac.jp}


\maketitle

\begin{abstract}
There are several lines of evidence that the super-massive black hole at
the Galactic center had higher activities in the past than directly
observed at present. Here I show that these lines of evidence can
quantitatively and consistently be explained if the mean accretion rate
during the past $\sim 10^7$ yrs has been $\sim 10^{3-4}$ times higher
than the current rate, by the picture of radiatively inefficient
accretion flow (RIAF) and associated outflow that has been successfully
applied to Sgr A$^*$. I argue that this increased rate and its duration
are theoretically reasonable in the Galactic center environment, while
the accretion rate suddenly dropped about 300 years ago most likely
because of the shell passage of the supernova remnant Sgr A East. The
chance probability of witnessing Sgr A$^*$ in such a low state is not
extremely small ($\sim$ 0.5\%). The outflow energetics is sufficient to
keep the hot ($\sim 8$ keV) diffuse gas observed in the Galactic center
region.  Then, I show that a significant amount of positrons should have
been created around the event horizon during the higher activity phase,
and injected into interstellar medium by the outflow. The predicted
positron production rate and propagation distance are close to those
required to explain the observed 511 keV annihilation line emission from
the Galactic bulge, giving a natural explanation for the large
bulge-to-disk ratio of the emission. The expected injection energy into
interstellar medium is $\sim$ MeV, which is also favorable as an
explanation of the 511 keV line emission.
\end{abstract}

\section{Introduction}

A variety of active and high energy phenomena are seen in the direction
towards the Galactic center over a broad range of wavelengths. It is
well established that the center of our Galaxy, Sagittarius (Sgr) A$^*$,
is a supermassive black hole (SMBH) with a mass of $M_\bullet \sim 3
\times 10^6 M_\odot$ (Genzel et al. 1997; Sch\"odel et al. 2003), and a
number of models have been proposed to explain the radiation and its
spectral energy distribution (SED) from Sgr A$^*$ by accretion of matter
onto it (see Baganoff et al. 2003 for a review and comparison to X-ray
observations). The concept of radiatively inefficient accretion flow
(RIAF) is theoretically well-motivated based on the physics of accretion
flows, and models based on the RIAF picture have successfully been
applied to many low accretion rate systems with $\dot M \lesssim 
0.1 \dot M_{\rm Edd}$, where $\dot M_{\rm Edd}$ is the Eddington mass accretion
rate (Narayan \& Quataert 2005 for a review).  Sgr A$^*$ is one of the
best studied examples, and the model of Yuan, Quataert, \& Narayan
(2003, 2004, hereafter YQN03 and YQN04, respectively) can reproduce the
observed SED of Sgr A$^*$ in a wide range of wavelengths from radio to
X-ray bands. An important ingredient of this model is that the accretion
rate decreases with decreasing radius from the SMBH, indicating a
significant mass loss by a magnetically driven outflow or wind. This is
necessary to make the model consistent with observations, and such wind
activities are also supported by recent numerical simulations (YQN03 and
references therein).

Though this model can explain the SED of Sgr A$^*$ at present, there are
several lines of evidence for higher activities in the Galactic center
in the past, such as much higher X-ray luminosity and mass outflows (see
\S \ref{section:past-activity} for a brief review).  Hence, it is
interesting to examine whether these lines of evidence can consistently
be explained by changing the accretion rate within the framework of the
RIAF model. The first aim of this paper is to show that it is indeed
possible, and I also discuss whether such a high accretion rate is
reasonable in the environment of the Galactic center, as well as the cause of
the sudden decrease leading to the current low rate.

The 511 keV electron-positron annihilation line emission in the Galaxy
has been observed for a long time (Kn\"odlseder et al. 2005 and
references therein), and the latest observation by the INTEGRAL
gamma-ray observatory revealed that the all sky distribution is
dominated by the extended bulge component ($\sim 8^\circ$ FWHM), with a
weak evidence for the disk component (Kn\"odlseder et al. 2005). Although
the disk component can be explained by positron emission from
radioactive nuclei produced in supernovae, the origin of the bulge
component is still a mystery.  A positron production rate of $\sim 1.5
\times 10^{43} \ \rm s^{-1}$ is required to explain the bulge component.

A number of candidates have been proposed for the origin of the bulge
component, but the large bulge-to-disk ratio excludes many of them
related to recent star formation activities, leaving type Ia supernovae
(SNe Ia) and low-mass X-ray binaries (LMXBs) as the primary candidates
(Kn\"odlseder et al. 2005). However, about one order of magnitude higher
rate of bulge SNe Ia than the best estimate is required to explain the
observed annihilation rate (Prantzos 2006). The positron production rate
from LMXBs or microquasars is highly uncertain and might be sufficient
(Guessoum et al. 2006), but the bulge-to-disk ratio of LMXBs is
considerably lower than that of the 511 keV data. Most of the positrons
produced in microquasars are probably lost by annihilation before
injection into interstellar medium (ISM) (Guessoum et al. 2006), and
direct annihilation gamma-ray emission from the sources would be
inconsistent with the diffuse gamma-ray background around MeV band (see
\S \ref{section:background-radiation}). Annihilation of dark matter
particles with a mass of $\sim$ MeV is another possible solution, but
such a particle is not naturally predicted by the particle physics
theory, in contrast to the well-motivated candidates such as the
supersymmetric particles that are much more massive ($m_\chi \gtrsim 50$
GeV, Bertone et al. 2004). Furthermore, the upper limit on 511 keV line
flux from the Sagittarius dwarf galaxy excludes this scenario for almost
all types of the halo density profile (Kn\"odlseder et al. 2005).

The second aim of this paper is to show that, if the past accretion rate
onto Sgr A$^*$ was in fact higher than now with the rate and duration
inferred from the observational evidence mentioned above, the RIAF model
of Sgr A$^*$ gives a natural explanation for the bulge 511 keV line
emission. There are models of positron production by accretion activity
of Sgr A$^*$ (Titarchuk \& Chardonnet 2006; Cheng, Chernyshov, \& Dogiel
2006), but the model proposed here is vastly different from these and it
gives a more natural explanation for the 511 keV line
(\S\ref{section:comparison}).

A variety of phenomena on various scales around the Galactic center will
be discussed in this paper, and these are summarized in schematic
diagrams in Fig. \ref{fig:schem} for the reader's convenience. The
summary of various comparison between observations and the model is
given in Table \ref{table:summary}.  Starting from a brief description
about the RIAF model of Sgr A$^*$ (\S\ref{section:RIAF}), I review and
summarize the lines of evidence for the past increased activity of Sgr
A$^*$, and present RIAF interpretations for these, in
\S\ref{section:past-activity}.  We discuss the physical origin of the
past high activity in the Galactic center environment in
\S\ref{section:origin}. Then I calculate the positron production rate
around the event horizon of Sgr A$^*$ in \S \ref{section:production},
and consider the ejection from Sgr A$^*$ (\S \ref{section:ejection}) and
propagation in ISM (\S \ref{section:propagation}).  In \S
\ref{section:discussion}, I compare my model with the other models of
the 511 keV emission, and discuss about future observational tests of
this model. I apply 8 kpc as the distance to the Galactic center
(Eisenhauer et al. 2003).

\section{The RIAF Model of Sgr A$^*$ and Outflow}
\label{section:RIAF}

The YQN03 model is a one-dimensional, global RIAF model that can
reproduce the SED of Sgr A$^*$. Though detailed three-dimensional
structure is not taken into account in this model, models or simulations
incorporating more complicated structures have not yet succeeded in
reproducing the Sgr A$^*$ SED (e.g., Mineshige, Kusnose, \& Matsumoto
1995; Ohsuga, Kato, \& Mineshige 2005).  In this paper I discuss the
scaling of the radiation and outflow from Sgr A$^*$ when the accretion
rate is changed, and the simple YQN03 model is sufficient and currently
the best choice for this purpose.

The accretion rate of the YQN03 model is normalized at the outer
boundary corresponding to the Bondi radius ($r_B \sim 10^5 r_s = 0.029$
pc), which is inferred from the ambient density and temperature of X-ray
emitting gas (Baganoff et al. 2003), where $r_s$ is the Schwartzshild
radius of the SMBH. The size of the extended ($\sim 1''.4$ FWHM) X-ray
emission of Sgr A$^*$ (Baganoff et al. 2003) nicely corresponds to this
radius. The accretion rate at this radius is $\dot M_{\rm acc}(r_B) \sim
\alpha_v \dot M_B \sim 10^{-6} M_\odot$/yr, where $\alpha_v \sim 0.1$ is
the dimensionless viscosity parameter and $\dot M_B$ is the Bondi
accretion rate. The YQN03 model assumes a radially varying accretion
rate, $\dot M_{\rm acc} \propto r^s$ with $s = 0.27$, based on the
adiabatic inflow-outflow solutions (ADIOS, Blandford \& Begelman 1999)
rather than the advection dominated accretion flow (ADAF, see Kato et
al. 1998; Narayan et al. 1998 for reviews) that is the simplest RIAF
solution with a radially constant mass accretion rate. This modification
from ADAF is necessary since the ADAF model predicts too large density
and magnetic field strength in the innermost region, in contradiction
with the observational upper bounds on rotation measure.

The accretion flow is quasi-spherical and the density 
$\rho$ is determined by
$\dot M_{\rm acc}(r) \sim 4 \pi \rho r^2 \upsilon_{\rm in}$, where
$\upsilon_{\rm in}$ is the inflow velocity of the accretion flow.  In
the self-similar solution of RIAFs, $\upsilon_{\rm in}$ is close to and
proportional to the free fall velocity $\upsilon_{\rm ff}$, and hence
$\rho \propto r^{-3/2+s}$. In the YQN03 model, the particle density is
$n_e \sim \rho/m_p \sim 3.9 \times 10^7 (r/r_s)^{-1.23} \rm \ cm^{-3}$, where
$m_p$ is the proton mass, and hence $\upsilon_{\rm in} \sim 0.15
\upsilon_{\rm ff}$. This also means $\rho \propto \dot M_{\rm acc}$.

It is reasonable to assume that the evaporated mass flow due to the
non-conserving accretion rate is ejected as wind or outflow, whose
velocity is comparable with the escape velocity $\upsilon_{\rm esc}$ at
the region where the wind originates. Then the mass ejection rate per
logarithmic interval of radius is $\dot M_w = r (d\dot M_{\rm acc}/dr) =
s \dot M_{\rm acc}$, and the largest kinetic energy is produced from the
innermost region where $\upsilon_{\rm esc}$ is close to the speed of
light, $c$. We denote $r \sim r_*$ for the wind production region, and
the outflow kinetic energy from this region becomes $L_{\rm kin} \sim
\dot M_w \upsilon_{\rm esc}^2 / 2 \sim 1.5 \times 10^{38} r_3^{-0.73}$
erg/s, where $r_3 \equiv r_* / (3 r_s)$.

The jet or outflow may also contribute to the Sgr A$^*$ SED, especially
in the radio bands (see discussion of YQN03). The outflow kinetic energy
derived here is in fact comparable with those assumed in the jet models
of Sgr A$^*$ (Falcke \& Biermann 1999; Yuan, Markoff, \& Falcke 2002; Le
\& Becker 2004), indicating that this estimate is a reasonable one. 

The outflow kinetic power is much larger than the X-ray luminosity of
Sgr A$^*$, which is $L_X \sim 10^{33}$ erg/s in the quiescent state and
$\sim 10^{34-35}$ erg/s in the flaring state.  Such a trend of
jet-energy dominance is indeed established for black holes in stellar
X-ray binaries in the low/hard state (Gallo et al. 2003, 2005), which is
believed to be a state corresponding to RIAF (Esin et al. 1997). In
fact, the above kinetic and X-ray luminosities are consistent with the
relation inferred for the stellar black hole binaries: $l_{\rm kin} = f
(l_X/0.02)^{0.5}$ with $f = 0.06$--1, where $l_{\rm kin}$ and $l_X$ are
in units of the Eddington luminosity (Gallo et al. 2005). Though the
mass scale of black holes is quite different, this is expected if (1)
the critical accretion rate in units of the Eddington rate ($\dot M /
\dot M_{\rm Edd}$) between the standard thin disk and RIAF is
independent of the black hole mass, (2) X-ray luminosity scales roughly
as $\dot M_{\rm acc}^2$ in the RIAF regime, and (3) the mass outflow is
proportional to $\dot M_{\rm acc}$. The first two are indeed the
properties of the ADAF (Narayan et al. 1998), and the third is in
accordance with the radially varying accretion rate assumed in YQN03.

Some active galactic nuclei (AGNs) show the jet activity with much
higher speed (the bulk Lorentz factor $\Gamma \sim 10$) than the above
velocity estimate ($\sim \upsilon_{\rm esc}$, only mildly relativistic
with $\Gamma \sim 1$). However, $\Gamma \sim 1$ is reasonable for Sgr
A$^*$, since the same trend has been known for the outflows from stellar
X-ray binaries in the low/hard state, which are not strongly beamed and
not extremely relativistic ($\Gamma \lesssim 2$), in contrast to those
of X-ray transients (Gallo et al. 2003; Narayan \& McClintock 2005).

\section{Past Higher Activity and RIAF Interpretation}
\label{section:past-activity}

\subsection{Higher X-ray Luminosity}
\label{section:past-X-ray-luminosity}

\begin{figure}
  \begin{center}
    \FigureFile(140mm,140mm){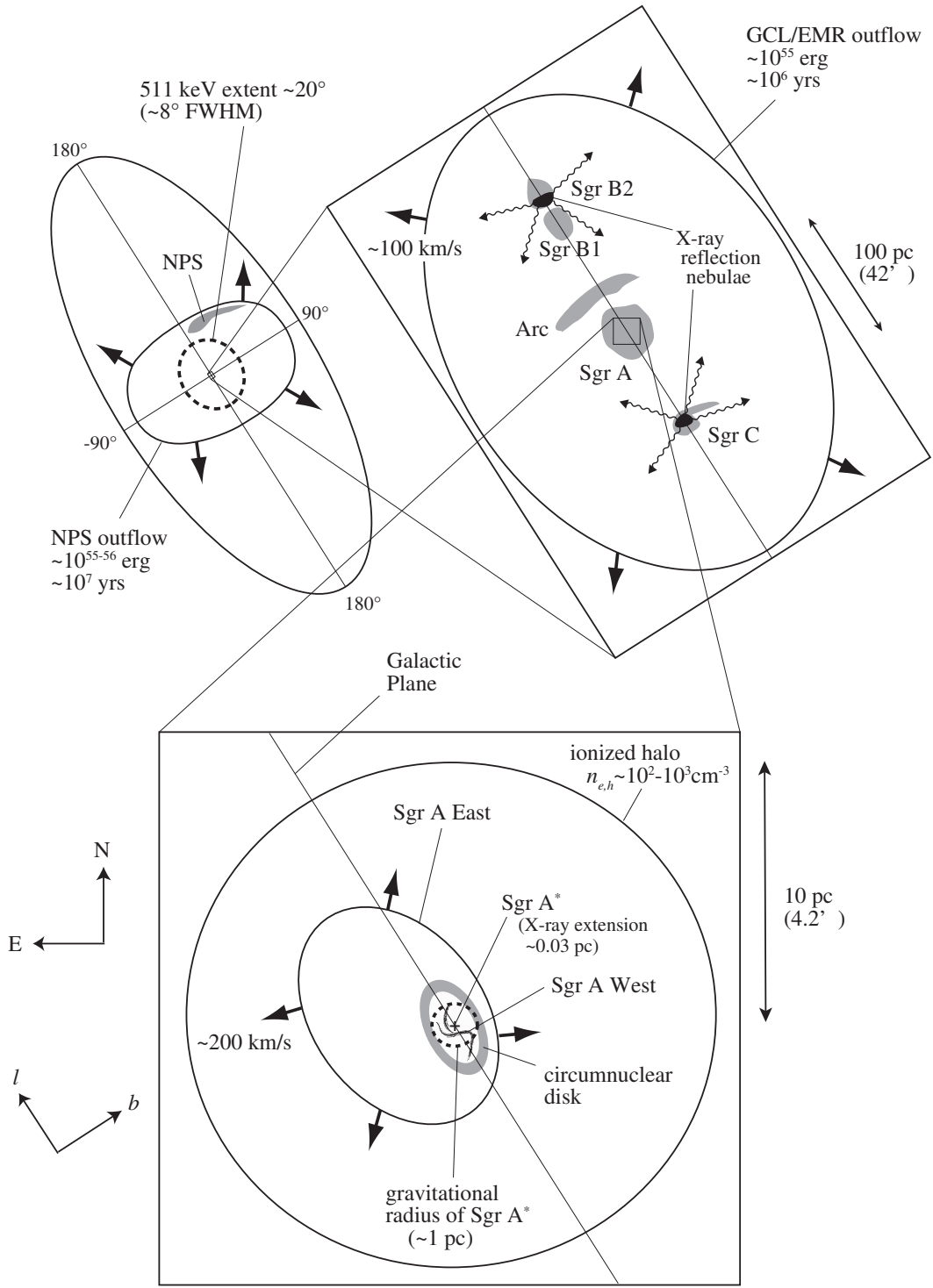}
  \end{center}
  \caption{ Schematic diagrams for various phenomena on various scales
 around the Galactic center discussed in this paper. The upper-left
 diagram is the all sky map showing the scales of the 511 keV
 annihilation line emission and the outflow making the North Polar Spur
 (NPS). The upper-right diagram is for the region within $\sim 1^\circ$
 of the Galactic center, showing the scale of the outflow observed as
 the Galactic center lobe (GCL) and the expanding molecular ring
 (EMR). Famous objects found in the radio image of this region are shown
 by greyscale, and the X-ray reflection nebulae are indicated by the
 black regions.  The lower diagram is for the innermost region, showing
 the interaction of Sgr A East, West, and Sgr A$^*$ surrounded by the
 ionized halo. The gravitational radius (dashed circle) is a radius
 within which the SMBH gravity is dominant compared with the surrounding
 stars.  } \label{fig:schem}
\end{figure}

There are a few independent lines of evidence that about 300 years ago
Sgr A$^*$ was much more luminous than now in the X-ray band. Koyama
et al. (1996) found fluorescent X-ray emission reflected from cold iron
atoms in the giant molecular cloud Sgr B2 by ASCA observation
(Fig. \ref{fig:schem}).  Since there is no irradiation source to explain
the iron line emission, they suggested a possibility that $\sim$300 yrs
ago Sgr A$^*$ was much brighter than now. More recent studies by
Murakami et al. (2000, 2001a, b) found a new X-ray reflection nebula
associated with Sgr C, and estimated the increased past luminosity of
Sgr A$^*$ as $L_X \sim 3 \times 10^{39}$ erg/s from Sgr B2 and C
data. This claim was independently confirmed by an INTEGRAL observation
covering higher energy band of 10--100 keV (Revnivtsev et al. 2004); the
ASCA and INTEGRAL data of Sgr B2 can nicely be fit by a reflection of
radiation from Sgr A$^*$, whose luminosity is $1.5 \times 10^{39}$ erg/s
in 2--200 keV and spectrum is a power-law with a photon index of $\beta
= 1.8 \pm 0.2$ ($dN/dE \propto E^{-\beta}$).

``The ionized halo'' surrounding Sgr A$^*$ with a density $n_{e, h} \sim
10^2$--$10^3 \ \rm cm^{-3}$ has been known from radio observations
(Pedlar et al. 1989; Anantharamaiah et al. 1999), and it extends to a
radius of $\sim 10$ pc (Fig. \ref{fig:schem}).  Maeda et al. (2002)
argued that currently no ionizing source is found for the ionized halo,
and the past activity of the Sgr A$^*$ may be responsible for the
ionization, requiring an X-ray luminosity of $L_X \sim 10^{40}$ erg/s, a
similar flux to those inferred from the X-ray reflection nebulae.

We can then estimate the boost factor of the accretion rate to achieve
the X-ray luminosity of $L_X \sim 10^{39}$--$10^{40}$ erg/s in the RIAF
model. It should be noted that this luminosity is $\sim 10^{-5}$ in
units of the Eddington luminosity, and hence it is still well within the
ADAF/RIAF regime. According to Fig. 5 of YQN04, a boost factor $f_b
\equiv \dot M_{\rm past} / \dot M_{\rm present} \sim 10^3$--$10^4$ (at a
fixed radius $r$) is required to achieve the luminosity of $L_X \sim
10^{39}$--$10^{40}$ erg/s, assuming a constant value for $s$. The SED in
YQN04 with $f_b \sim 10^3$ is roughly constant in $\nu L_\nu$
(luminosity per logarithmic frequency interval) in the X-ray band, being
consistent with the X-ray spectrum inferred from the X-ray reflection
nebula. Then the outflow kinetic luminosity in such a higher activity
phase should be $L_{\rm kin} \sim 4.7 \times 10^{41} f_{3.5}
r_3^{-0.73}$ erg/s, where $f_{3.5} \equiv f_b / 10^{3.5}$.  In units of
the Eddington mass accretion rate, $\dot M_{\rm Edd} = 10 L_{\rm
Edd}/c^2$ (corresponding to a radiative efficiency of 0.1), the
accretion rates at $r = r_B$ and $r = r_s$ are $4.6 \times 10^{-2}
f_{3.5}$ and $2.1 \times 10^{-3} f_{3.5}$, respectively.

\subsection{Mass Outflows on Various Scales} 
\label{section:outflow}

There is evidence for powerful mass outflow from the Galactic center on
scales from arcminutes to tens of degrees (Fig. \ref{fig:schem}).
Bland-Hawthorn \& Cohen (2003) reported the mid-infrared emission from
dust expanding in the Galactic center lobe (GCL) on a few degree scale,
and estimated the total kinetic energy as $\sim 10^{55}$ erg with a
velocity of $\sim$ 100 km/s and a dynamical time of $\sim 10^6$ yr. The
size, energy, and time scales are similar to those of the expanding
molecular ring (EMR) around the Galactic center (Kaifu et al. 1972;
Scoville 1972). Bland-Hawthorn \& Cohen further argued that this result
is consistent with the North Polar Spur (NPS) on an even larger scale
(up to tens of degrees), which has been interpreted by Sofue (2000) to
be an outflow from the Galactic center \footnote{There are other
interpretations for NPS by closer objects on the Galactic disk like a
supernova remnant, but see Bland-Hawthorn \& Cohen (2003) for arguments
favoring the Galactic center interpretation.}, with an energy scale of
$\sim 10^{55-56}$ erg and a dynamical time scale of $\sim 10^7$ yr.  The
kinetic luminosity inferred from these various observations is in a nice
agreement with the estimate for the RIAF model with the boost factor of
$f_b \sim 10^{3-4}$, and hence we can attribute these outflows to the
past activity of Sgr A$^*$ that was responsible for the higher X-ray
luminosity. This indicates the duration of $\sim 10^7$ yr for the higher
activity, which is reasonable compared with various estimates for
lifetimes of AGNs (Martini 2004).

The origin of the outflow from the Galactic center may instead be
starbursts in the nuclear region, as suggested by Bland-Hawthorn \&
Cohen (2003).  However, observations of gamma-rays from radioactive
$^{26}$Al seem to disfavor the starburst scenario. The energy of
$10^{55}$ erg in $10^6$ yr is equivalent to $\sim 1$ supernova per
century, almost comparable with the Galaxy-wide core-collapse supernova
rate of $1.9 \pm 1.1$ per century (Diehl et al. 2006) estimated from the
flux of $^{26}$Al gamma-rays, whose spatial distribution is clearly
associated along with the Galactic disk. Since the half-life of
$^{26}$Al is $7.2 \times 10^5$ yr, $^{26}$Al cannot travel beyond the
GCL region with the outflow velocity of $\sim 100$ km/s.
Therefore, if the origin of the outflow is starbursts, we expect a
strong $^{26}$Al gamma-ray emission concentrated within a few degree
from the Galactic center, with a flux comparable with the total
gamma-ray flux along with the disk. However, such a strong concentration
at the Galactic center is not found (Prantzos \& Diehl 1996;
Kn\"odlseder et al. 1999), indicating that the accretion activity of Sgr
A$^*$ is more plausible as the origin of the mass outflow.

Muno et al. (2004) studied in detail the Chandra data of the diffuse
X-ray emission within $\sim $ 20 pc of the Galactic center, and they
concluded that the hard component plasma ($k T \sim 8$ keV, Koyama et
al. 1989; Yamauchi et al. 1990) cannot be explained by unresolved point
sources. The latest Suzaku observation by Koyama et al. (2006)
further strengthened the case for the truly diffuse plasma
(but see also Revnivtsev et al. 2006). If it is truly diffuse,
it cannot be gravitationally bound and requires a large energy input of
$\sim 10^{40}$ erg/s to keep this hot plasma, when the escape time is
estimated simply by the sound velocity.  This is too large to be
explained by supernova explosions and the origin of this hot plasma is
still a matter of debate. Later in this paper (\S
\ref{section:wind-dynamics}), I will discuss in more detail about the
consequences of the energy injection into ISM by the wind from Sgr
A$^*$, and show that this hot plasma can be explained as a result of
shock heating by the wind.

\section{The Origin of the Past Higher Activity}
\label{section:origin}

We have shown that all the lines of evidence for the past higher
activities of Sgr A$^*$ can be explained if the mean accretion rate in
the past $\sim 10^7$ yrs is higher than now by a factor of $f_b \sim
10^{3-4}$, and Sgr A$^*$ had such a high rate 300 yrs ago. Then the next
questions are: (1) what is the source of the accreting matter during the
high activity phase, and (2) what caused the sudden drop of accretion
rate by a factor of $10^{3-4}$ on a time scale of just $\sim 10^{2-3}$
yrs. Here I give reasonable explanations for these.

\subsection{The Role of Sgr A East and the Ionized Halo}

Maeda et al. (2002) proposed that, based on their Chandra observation of
the supernova remnant Sgr A East, the past higher activity was induced by
accretion from the dense supernova shell expanding into the ionized
halo.  The location of Sgr A$^*$ is in fact inside Sgr A East and close
to its shell (Fig. \ref{fig:schem}).  In this scenario the duration of
such high accretion rate is only $\sim 10^3$ yr, as inferred from the
shell thickness ($\sim$1/10 of the observed shell radius $r_{\rm sh} = $
2.9 pc) and the shell expansion velocity ($\upsilon_{\rm sh} \sim$ 200
km/s estimated by a simple theoretical model of supernova remnants).
For comparison, the age estimate of Sgr A East is $\sim 10^4$ yr. A much
higher accretion rate than now is possible by the Bondi accretion with 
the shell density estimated from shock-compression of the gas of the
ionized halo. (The supernova ejecta is negligible at this stage.)

However, according to this scenario, we expect a comparable or even
higher accretion rate by accretion directly from the ionized halo before
the passage of the Sgr A East shell, since the density enhancement by
shock-compression is at most by a factor of 4 and the sound velocity of
unshocked gas is likely much lower than the shell velocity. Suppose that
the SMBH is embedded directly in the ionized halo. The Bondi radius in
this case could be different from the current estimate (0.03 pc) based
on the X-ray observation (Baganoff et al. 2003), since the current gas
properties around Sgr A$^*$ have been altered from those of the ionized
gas by the passage of the Sgr A East shell.  The Bondi radius in the
halo, $r_{B, h} = 2 GM_\bullet/c_{s, h}^2$ is larger than 1 pc if the
temperature $k T_h$ is lower than $\sim$ 110 eV, where $c_{s, h}$ is
the sound velocity in the ionized halo. A plausible temperature of the
ionized halo is $\sim 1$ eV (Maeda et al. 2002), and hence the outer
boundary of the accretion flow would not be determined by the Bondi
radius, but rather by the gravitational radius $r_{\rm grav} \sim 1$ pc,
within which the gravity of the SMBH is dominant compared with stars
around the SMBH (e.g., Sch\"odel et al. 2003).

Then we can extrapolate the YQN03 model from $r_B = 0.029$ pc out to
$r_{\rm grav}$ with the boost factor of $f_b \sim 10^{3-4}$ and $n_e
\propto r^{-1.23}$, and the density at $r_{\rm grav}$ becomes $n_e \sim
1.1 \times 10^3 f_{3.5} \ \rm cm^{-3}$. This is consistent with the
density of the ionized halo, $n_{e, h}$, and hence the RIAF at the
increased accretion rate is naturally connected to the environment
surrounding the SMBH.  Since this is an extrapolation of the RIAF
solution, the surrounding gas can accrete even if it has a significant
angular momentum, because of the angular momentum loss by
viscosity. This is in contrast to the simple picture of the spherical
Bondi accretion.

In this new scenario, the high accretion rate can last for a much longer
time scale than $\sim 10^3$ yrs, and accretion from the dense supernova
shell is no longer necessary. Still, Sgr A East must play an important
role to explain the sudden drop of the accretion rate $\sim$300 yrs ago,
by the destruction of accretion flow when the dense shell passed through
the SMBH. It should also be noted that there are the arm-like structures
of Sgr A West and the circumnuclear disk surrounding it on the scale of
$\sim$1 pc (e.g., Yusef-Zadeh, Melia, \& Wardle 2000; see also
Fig. \ref{fig:schem}), which could be the remnants of the former
accretion flow.

\subsection{Destruction of the Accretion Flow by Sgr A East}

We examine the destruction process more quantitatively as
follows. Since the ADAF and ADIOS solutions have a positive Bernoulli
parameter (Narayan et al. 1998; Blandford \& Begelman 1999), the flow is
not gravitationally bound and change of flow velocity by external force
would result in a destruction of the flow.  Therefore we should compare
the momentum of the flow and the supernova remnant to estimate the
effect of the supernova shell passage.  The momentum of the accretion
flow is:
\begin{eqnarray}
P_{\rm acc} &\sim& \frac{4\pi}{3} r^3 \rho \ \upsilon_{\rm flow} \\
&\sim& 5.6 \times 10^{41} \left(\frac{r}{1 \ \rm pc}\right)^{1.27} 
f_{3.5} \ \rm g \ cm \ s^{-1},
\end{eqnarray}
where we estimated the flow velocity $\upsilon_{\rm flow}$ by
$\sim \upsilon_{\rm in}$, since the rotation velocity
is negligible when the adiabatic index $\gamma_{\rm ad} \rightarrow
5/3$ in ADAFs (Narayan et al. 1998).
The momentum of the supernova remnant given to the flow is:
\begin{eqnarray}
P_{\rm SN} &=& \frac{1}{4}\left(\frac{r}{r_{\rm sh}}\right)^2 
\frac{2 E_{\rm SN}}{\upsilon_{\rm sh}} \\
&\sim& 3.0 \times 10^{42} \left(\frac{r}{1 \ \rm pc}\right)^2 
\left(\frac{E_{\rm SN}}{10^{51} \ \rm erg}\right)
 \ \rm g \ cm \ s^{-1} \ ,
\end{eqnarray}
where $E_{\rm SN}$ is the shell kinetic energy of the supernova remnant.
Therefore the accretion flow could have been destroyed at $r \gtrsim
0.1$ pc from the SMBH. Destruction should have
occurred with a time scale of the shell
crossing ($\sim 10^3$ yrs), and the accretion time scale at this radius
is also $r/\upsilon_{\rm in} \sim 1.3 \times 10^3$ yrs.  These time
scales are consistent with the required time scale of 
the accretion rate drop, $\sim 300$ yrs.  The
current low rate may be determined by either the diffuse gas in the
supernova remnant, winds from nearby stars (e.g., Cuadra et al. 2006), or
residual of the former accretion flow.

\subsection{Comparison with Nearby Galaxies}

We may ask how the suggested high activity of Sgr A$^*$ in the past
compares with those found in nearby normal galaxies, because it would be
statistically unlikely if our Galaxy had a much higher activity compared
with nearby normal galaxies. As reviewed by Ho (2004), nuclear activity
is quite commonly found in nearby galaxies. The YQN model with $f_b
\gtrsim 10^3$ predicts a nuclear $B$-band luminosity of $L_B \sim
10^{39}$ erg/s, and the number density of such galaxies expected from
the nuclear luminosity function is $\sim 10^{-2} (h/0.75)^3 \ \rm
Mpc^{-3}$, which is similar to that of typical galaxies like our own,
where $h = H_0/(\rm 100 \ km/s/Mpc)$ is the Hubble constant.  Nuclear
radio cores with flux of $10^{19}$--$10^{21} \ \rm W \ Hz^{-1}$ at 5 GHz
are also commonly found, which is again a similar radio luminosity to
that predicted by the YQN04 model with $f_b \sim 10^{3-4}$. Therefore,
the increased activity of Sgr A* is not particularly rare compared with
nearby normal galaxies, indicating that the characteristic time scale of
the increased activity can be $\sim 10^7$ yr or even longer, possibly as
long as the cosmological time scale. The accretion rate at the event
horizon, i.e., the mass growth rate of the SMBH is $1.4 \times 10^{-4}
f_{3.5} M_\odot \rm yr^{-1}$, and the mass growth in 10 Gyr is $1.4
\times 10^6 f_{3.5} M_\odot$, which is still less than $M_\bullet$.

AGN activity is generally sporadic and showing strong variability, and
hence it is naturally expected that the accretion rate was changing
significantly always in the past $\sim 10^7$ yrs. In fact, the
characteristic structures such as GCL, EMR, and NPS indicate such
variability or modulation of the accretion and wind activity. However,
conclusions of this paper are not seriously affected if the mean or
characteristic accretion rate is given by $f_b \sim 10^{3-4}$.

\subsection{Are We Living in a Very Special Time?}

One may feel uneasy if we are witnessing a large drop of the accretion
rate just during $\sim$300 yrs compared with the normal accretion time
scale of $\sim 10^7$ yrs, indicating an extremely low chance probability
of observing Sgr A$^*$ in the present phase: $300/10^7 = 3 \times
10^{-5}$. Here I argue that the actual chance probability is likely to
be much larger than this simple estimate.

About 100 young massive stars are known within the central parsec of Sgr
A$^*$ (Paumard et al. 2006). Some of them are already in the
post-main-sequence phase, and the age of this stellar population is
estimated as $\sim 6$ Myr.  This indicates a supernova rate of $\sim 1.7
\times 10^{-5} \rm yr^{-1}$, and the chance probability of observing Sgr
A$^*$ within 300 yrs after the shell passage becomes $0.5 \%$. This is
small, but not extraordinarily small for realization.
The mean accretion rate can be kept at the level of $f_b \sim 10^{3-4}$
if the accretion rate comes back to the mean level within the typical
time interval of supernovae, $\sim 6 \times 10^4$ yrs, after the shell
passage.  

It is an observational {\it fact} that we are living in a somewhat
special time, if the association of Sgr A$^*$ with the Sgr A East shell
is not just a superposition in the sky but physical. The age of $\sim
10^4$ yrs of Sgr A East indicates that the supernova rate within the
central parsec of Sgr A$^*$ should be less than $10^{-4} \ \rm
yr^{-1}$. Therefore, the chance probability must be less than $300/10^4
= 3\%$, which is just 6 times larger than the above estimate. Therefore
I consider that the argument of the chance probability does not
seriously weaken the case for the overall picture proposed by this work.

\section{Positron Production around Sgr A$^*$}
\label{section:production}
\subsection{Physical Quantities around the Event Horizon}
\label{section:quantities}

Now I consider the pair-production at the region where the wind
originates, $r \sim r_*$.  During the increased phase, the particle
accretion rate is $\dot N_{\rm acc} = \dot M_{\rm acc}/m_p \sim 7.1
\times 10^{45} f_{3.5} r_3^{0.27} \ \rm s^{-1}$ and the particle outflow rate
per logarithmic radius is $\dot N_w = s \dot N_{\rm acc}$. 
The particle density in the accretion flow is $n_e \sim 3.2 \times
10^{10} f_{3.5} r_3^{-1.23} \rm \ cm^{-3}$.
The accretion time spent around this radius is
$t_{\rm acc} \sim r_*/\upsilon_{\rm in}\sim 1.0 \times 10^3 r_3^{3/2}$ s.

The electron temperature of the YQN03 model is $T_e \sim 8 \times
10^{10} r_3^{-1}$ K, and hence relative motion of electrons is
sufficiently relativistic at the pair-production region if $r_* \sim 3
r_s$, with the mean electron Lorentz factor $\gamma_{e} \sim 3.151 k T_e
/ (m_e c^2) \sim 40 r_3^{-1}$ in the rest frame of the accretion
flow. The temperature does not heavily depend on the enhancement factor
$f_b \sim 10^{3.5}$, if the transfer efficiency of viscous heating
energy from ions to electrons (the parameter $\delta$ in YQN03) does not
change with the accretion rate, as assumed by YQN04. It may increase with
increasing accretion rate because of higher density and hence more
efficient interactions, but the value of $\delta$ assumed in YQN03 is
already of order unity (=0.55), not leaving much room for the increase
of $\delta$.

\subsection{Pair Equilibrium Criterion}
\label{section:equilibrium}

The positron density produced in the accretion flow depends on whether
the $e^\pm$ pair production process achieves the equilibrium with the
$e^\pm$ pair annihilation. This can be evaluated by comparing the pair
production rate density $\dot n_+$ and pair annihilation rate density
$\dot n_{\pm, \rm ann} \equiv n_e n_+ \sigma_{\pm, \rm ann} c$, 
where $n_+$ is the
produced positron density. The $e^\pm$ pair annihilation 
cross section is
\begin{eqnarray}
\sigma_{\pm, \rm ann} &=& \frac{\pi r_e^2}{\gamma_{\pm}} [\ln 2 \gamma_{\pm}
- 1]  
\end{eqnarray}
in the ultra-relativistic limit (Svensson 1982), where $r_e$ is the
classical electron radius and $\gamma_\pm$ is the Lorentz factor in the
rest frame of one particle.
Hence $\gamma_\pm$ can be related as $\gamma_\pm \sim \gamma_+
\gamma_e$, where $\gamma_+$ is the positron Lorentz factor
in the flow frame. 
Initially the positron density is small, and
it increases with time as $n_+ \sim \dot n_+ t$ until $t \sim t_{\pm,
\rm ann}$ 
when the equilibrium is achieved ($\dot n_+ = \dot n_{\pm, \rm ann}$),
where the pair annihilation time scale is $t_{\pm, \rm ann} \equiv (n_e c 
\sigma_{\pm, \rm ann})^{-1}$.
Since $t_{\pm, \rm ann}$ does not depend on $\dot n_+$, the
equilibrium condition is the same for any pair production processes and
it is determined by comparing $t_{\pm, \rm ann}$ to the accretion time scale,
as:
\begin{eqnarray}
\frac{t_{\rm acc}}{t_{\pm, \rm ann}} &=& 4.2 \times 10^{-2} 
\gamma_+^{-1} f_{3.5} r_3^{1.27} \ ,
\end{eqnarray}
where we estimated the logarithmic part of the cross section by
$\gamma_+ \sim \gamma_e \sim 40$. 

For positrons produced by electron-electron scattering ($e^- e^-
\rightarrow e^- e^- e^- e^+$), we expect $\gamma_+ \sim \gamma_e$, while
for positrons produced by two photon annihilation ($\gamma \gamma
\rightarrow e^- e^+$), a variety of $\gamma_+$ is possible (see the
following subsections).  The positron energy may also significantly
change by interaction with the accreting material within the accretion
time scale. However, for any value of $\gamma_+$, we find that $t_{\rm
acc} \ll t_{\pm, \rm ann}$, and hence the equilibrium will not be
achieved, meaning that we can estimate the pair density by $n_+ \sim
\dot n_+ t_{\rm acc}$.

\subsection{Spectral Energy Distribution of Sgr A$^*$}

For pair production processes including photons, we must assume the form
of the SED and the emission region of Sgr A$^*$ during the increased
activity phase. Here I assume that the radiation is mainly coming from
the region around the event horizon. It should be noted that the X-ray
emission in the quiescent phase of the YQN03 model for present-day Sgr
A$^*$ is dominated by thermal bremsstrahlung at large radii far from the
SMBH, which is in agreement with the extended X-ray emission ($\sim
1''$). However, the synchrotron-self-Compton (SSC) component
becomes dominant when the accretion rate is increased as $f_b \gtrsim
10^3$ (YQN04), and it is produced in the innermost region.  Therefore the
above assumption is reasonable for the higher activity phase.

For calculations below, I assume a constant SED in $\nu L_\nu$. The
predicted SED of the YQN04 model when $f_b \gtrsim 10^3$ is
approximately flat in $\nu L_\nu$ from keV to MeV band.  A flat $\nu
L_\nu$ SED is also supported in 2--200 keV band by the observed spectrum
of the X-ray reflection nebula (\S \ref{section:past-X-ray-luminosity}).
Therefore, this assumption is well supported both by observation
and theory in the photon energy band of keV--MeV.

The SED beyond MeV is more uncertain, and here I simply examine the
constraints on the current SED of Sgr A$^*$.  The TeV gamma-ray emission
detected by H.E.S.S. from the inner $10'$ of Sgr A$^*$ (Aharonian et
al. 2004) is possibly coming from the region close to the event horizon,
and its flux is similar to that extrapolated from X-ray bands with a
constant $\nu L_\nu$ spectrum. Though the GeV flux of 3EG J1746$-$2851,
which is the closest to Sgr A$^*$ among the EGRET sources, is
considerably higher than the X-ray flux of Sgr A$^*$, the poor angular
resolution in this band does not allow to establish a firm connection
between the GeV flux and Sgr A$^*$ (Mayer-Hasselwander et al. 1998;
Aharonian \& Neronov 2005)\footnote{A recent analysis of the EGRET data
by Hooper \& Dingus (2004) excluded Sgr A$^*$ as the origin of 3EG
J1746$-$2851 beyond the 99.9\% confidence level.}.  Therefore, it is not
unreasonable to assume a flat $\nu L_\nu$ SED at photon energy of
$\gtrsim$ MeV, though uncertainty is large. On the other hand, the
photon production region will become optically thick to $e^\pm$ pair
production for very high energy photons ($\gtrsim$ GeV), and the
constant $\nu L_\nu$ assumption will be no longer valid at such high
energy band (see \S \ref{section:photon-photon}).

\subsection{Expected Pair Amount in the Wind}

Now I estimate the expected amount of pairs produced by three processes
of pair production, i.e., electron-electron scattering ($e^- e^-
\rightarrow e^- e^- e^- e^+$), photon-electron collisions ($\gamma e^-
\rightarrow e^- e^- e^+$), and two photon annihilation ($\gamma \gamma
\rightarrow e^\pm$). The rates of corresponding proton processes (e.g.,
$ p \ e^- \rightarrow p \ e^- e^- e^+$) are about one order of magnitude
smaller than these (Zdziarski 1982, 1985).

\vspace{0.2cm}

\subsubsection{Electron-Electron Scattering}
\label{section:electron-electron}

I used the formula given in Svensson (1982) for the cross section
in the ultra-relativistic limit, which
is $\sigma_{ee} = 1.7 \times 10^{-28} \ \rm cm^2$ for $\gamma_e = 40$
and depends on $\gamma_e$ only logarithmically. Hence I ignore the
dependence on $\gamma_e$. Then the density ratio of the produced
positrons to electrons is given as:
\begin{eqnarray}
\frac{n_+}{n_e} &=& \frac{\dot n_+ t_{\rm acc}}{n_e}
= c \sigma_{ee} n_e t_{\rm acc} \\
  &=& 1.6 \times 10^{-4} f_{3.5} r_3^{0.27}\ ,
\end{eqnarray}
and hence the total positron production rate as an outflow from
Sgr A$^*$ is:
\begin{eqnarray}
\dot N_+ &\sim& \dot N_w \left(\frac{n_+}{n_e}\right) \\
 &=& 3.2 \times 10^{41} f_{3.5}^2 r_3^{0.54} \ \rm s^{-1}  \ .
\end{eqnarray}
This is a rate per $\ln r_*$, and integrating from $r_* = 3 r_s$ to $40
\times 3 r_s$, beyond which electrons become non-relativistic and hence
the above formulations are no longer valid, the rate is increased by a
factor of 11.7 leading to $\dot N_+ \sim 3.7 \times 10^{42} f_{3.5}^2 \
\rm s^{-1}$.  This estimate is, within the model uncertainties, in nice
agreement with the rate required for the bulge 511 keV emission, $1.5
\times 10^{43} \ \rm s^{-1}$ (Kn\"odlseder et al. 2005).

\vspace{0.2cm}
\subsubsection{Photon-Electron Collisions}
\label{section:electron-photon}

The cross section for $e^-\gamma \rightarrow e^- e^- e^+$ depends on the
photon frequency $\nu_{\rm er}$ 
measured in the electron's rest frame, which
is given by (Svensson 1982)\footnote{The numerical factor 
$3 \sqrt{\pi} $ of the non-relativistic formula
in Svensson (1982, eq. 31) should be corrected
to $\pi \sqrt{3}$ as noted in Svensson (1984).}:
\begin{eqnarray}
\sigma_{e\gamma} = \frac{\pi \sqrt{3}}{972} 
\alpha r_e^2 \left( x_{\rm er} - 4 \right)^2
\end{eqnarray}
in the non-relativistic limit ($x_{\rm er}
- 4 \ll 4 $), and
\begin{eqnarray}
\sigma_{e\gamma} = \alpha r_e^2 \left[\frac{28}{9} \ln \left( 2 x_{\rm er}
\right) - \frac{218}{27} \right]
\end{eqnarray}
in the ultra-relativistic limit ($x_{\rm er} - 4 \gg 4$), where $x_{\rm er}
\equiv h\nu_{\rm er}/(m_e c^2)$, $\alpha$ is the fine structure
constant, and the reaction threshold is $x_{\rm er, th} = 4$.

Treating electrons as a single energy population with $\gamma_e \sim
40$, and estimating the photon number density per unit photon frequency
($\nu$) in the laboratory frame as $n_\nu \sim L_\nu / (4 \pi r_*^2 c h
\nu)$, the positron production rate by this process can be written as:
\begin{eqnarray}
\dot n_+ &=& \int n_e n_\nu \sigma_{e\gamma} c d\nu \\
&=& \frac{\gamma_e n_e (\nu L_\nu)}{4 \pi r^2 m_e c^2}
\int \frac{\sigma_{e\gamma}(x_{\rm er})}{x_{\rm er}^2} 
 dx_{\rm er} \ ,
\end{eqnarray}
where we have used $\nu_{\rm er} \sim \gamma_e \nu$.  With the
assumption of a flat $\nu L_\nu$ SED, the integration over $x_{\rm er}$
is mostly contributed from photons with $x_{\rm er} \sim 20$, i.e., $h
\nu \sim 0.3$ MeV for $\gamma_e \sim 40$. This target photon energy is
within the range where the Sgr A$^*$ luminosity during the increased
activity can reliably be assumed, and hence the uncertainty about the
luminosity and SED is small.

Then the produced positron density is given by:
\begin{eqnarray}
\frac{n_+}{n_e} &=& \frac{\dot n_+ t_{\rm acc}}{n_e} \sim
4.2 \times 10^{-4} L_{39.5} r_3^{-1.5} \ ,
\end{eqnarray}
where $L_{39.5} = \nu L_\nu / (10^{39.5} \ \rm erg/s)$, 
and the total positron production rate in the outflow is:
\begin{eqnarray}
\dot N_+ &\sim& \dot N_w \left(\frac{n_+}{n_e}\right) \\
 &=& 8.1 \times 10^{41} L_{39.5} f_{3.5} r_3^{-1.23}\ \rm s^{-1} \ .
\end{eqnarray}
Integration over $\ln r_*$ at $r_* > 3 r_s$ would slightly decrease the
above number by a factor of 1/1.23.  Though this number is smaller than
the rate required to explain the 511 keV emission by about one order of
magnitude, it may also be important if the model uncertainties are
considered. Note that $L_\nu \propto \dot M_{\rm acc}^2$ 
at photon energies higher than the X-ray band in RIAFs, and
hence $\dot N_+ \propto f_b^3$.

\vspace{0.2cm}
\subsubsection{Photon-Photon Annihilation}
\label{section:photon-photon}

The pair-production by two photon annihilation most efficiently occurs
with a cross section of $\sigma_{\gamma\gamma} \sim 1.7 \times 10^{-25}
\ \rm cm^{-2}$, when the photon energy at the center-of-mass is about
the electron mass, as $(h \nu_l h \nu_h)^{1/2} \sim 2 m_e c^2$, where
$\nu_l$ and $\nu_h$ are the frequencies of two photons at the laboratory
frame ($\nu_l < \nu_h$) (e.g., Salamon \& Stecker 1998).  For photons meeting
this condition, the pair-production rate density is given by
\begin{eqnarray}
\dot n_+ \sim
\nu_h n_\nu(\nu_h) \ \nu_l n_\nu(\nu_l) \ \sigma_{\gamma \gamma} c \ ,
\end{eqnarray}
which is constant against $\nu_h$ (and $\nu_l$) by the assumption
of the constant $\nu L_\nu$ SED. 
It should be noted that this is valid only
when the region is optically thin for high frequency photons, i.e.,
$\tau_{\gamma \gamma} \lesssim 1$, where
\begin{eqnarray}
\tau_{\gamma \gamma} &=& \nu_l n_\nu(\nu_l) \sigma_{\gamma
\gamma} r_* \\
&=& 3.4 \times 10^{-4}
L_{39.5} r_3^{-1} \left( \frac{h\nu_l}{\rm 1 \ MeV}
\right)^{-1} \ .
\end{eqnarray}
The luminosity and pair production would then be suppressed for very
high energy photons of $\gtrsim$ GeV.\footnote{Because of the increased
luminosity compared with that of Sgr A$^*$ at present, this energy scale
is much smaller than the estimate by Aharonian \& Neronov (2005) 
for the present-day Sgr A$^*$ ($\sim$
10 TeV).}  Therefore we expect that the pair production rate will mostly
be contributed by photons in keV--GeV bands.

Now the density of positrons produced is:
\begin{eqnarray}
\frac{n_+}{n_e} &=& \frac{\dot n_+ t_{\rm acc}}{n_e} 
\\
&=& 8.4 \times 10^{-5} L_{39.5}^2 f_{3.5}^{-1} r_3^{-1.27} \ ,
\end{eqnarray}
and hence the total positron production rate from the Sgr A$^*$ is:
\begin{eqnarray}
\dot N_+ &=& \dot N_w \left(\frac{n_+}{n_e}\right) \\
&=& 1.6 \times 10^{41} L_{39.5}^2 r_3^{-1} \ \rm s^{-1} \ .
\end{eqnarray}
Note that this estimate is per unit logarithmic interval of $\nu_h$ (or
$\nu_l$). Since the Sgr A$^*$ luminosity during the increased activity
can reliably be modeled up to $\sim$ MeV, the uncertainty is small for
photons of $h\nu_l \sim h\nu_h \sim m_e c^2$.  Though it suffers larger
uncertainty about the SED in MeV--GeV bands, integration over $h\nu_h
\sim $ 1 MeV--1 GeV would increase the total rate by a factor of
6.9. The rate has a large dependence on the accretion rate as $\dot N_+
\propto L_\nu^2 \propto f_b^4$.  Considering the model uncertainties,
the two photon annihilation may also substantially contribute to the
observed bulge 511 keV emission.

\subsection{Comparison with Previous Studies}

Positron production in RIAFs has been investigated in several previous
studies (Kusunose \& Mineshige 1996; Bj\"ornsson et al. 1996; Esin et
al. 1997), and the results presented here are consistent with these in a
sense that the effect of pair production in RIAFs is small, i.e., $n_+
\ll n_e$. However, the values of $(n_+/n_e)$ found in this work are
quantitatively higher than the results of Esin et al. (1997) for stellar
X-ray binaries, who found that the pair annihilation is in equilibrium
with the pair production by the electron-electron collisions, with the
density ratio of $n_+/n_e \lesssim 10^{-5}$. The difference can be
understood as follows.

As discussed in \S \ref{section:equilibrium}, 
the criterion of the equilibrium between pair production and
annihilation is given by $t_{\rm acc}/t_{\rm \pm, ann}$. 
According to the RIAF theory, I find the dependence of this quantity
on the relevant parameters as:
\begin{eqnarray}
\frac{t_{\rm acc}}{t_{\rm \pm, ann}} \propto \sigma_{\rm \pm, ann}
\ \frac{\dot M}{M_\bullet} \propto \sigma_{\rm \pm, ann}
\ \frac{\dot M}{\dot M_{\rm Edd}}   \ .
\end{eqnarray}
Though the Eddington accretion ratio $\dot M / \dot M_{\rm Edd}$ considered in
this work is within the parameter range investigated by Esin et
al. (1997), $\sigma_{\rm \pm, ann}$ is considerably different by the
difference of the electron temperature. In the two-temperature ADAF
model considered by Esin et al., the electron temperature is about the
virial value as $T_e \sim 10^{9-10}$ K, while it is much higher in the
RIAF model of Sgr A$^*$, as $T_e \sim 10^{11}$ K, because of the high
value of $\delta$ (\S \ref{section:quantities}).  Since $\sigma_{\pm,
\rm ann} \propto \gamma_\pm^{-1} \sim \gamma_e^{-2}$, the value of
$t_{\rm acc}/t_{\rm \pm, ann}$ for ADAF is much higher than the model
presented here, leading to the pair equilibrium 
($t_{\rm acc} > t_{\rm \pm, ann}$) as found by Esin et al.

In the pair equilibrium, the positron density is given by
$n_+ \sim \dot n_+ / (n_e \sigma_{\pm, \rm ann} c)$. 
In the case of the pair production by electron-electron collisions
($\dot n_+ \sim n_e^2 \sigma_{ee} c$), we find:
\begin{eqnarray}
\left(\frac{n_+}{n_e}\right)_{\rm eq} 
\sim \frac{\sigma_{ee}}{\sigma_{\pm, \rm ann}} \ .
\end{eqnarray}
In the model presented here, this value is $n_+/n_e \sim 0.15$ for
$\gamma_e \sim 40$, and hence $n_+/n_e > 10^{-5}$ is possible even if
the pair equilibrium is not reached. On the other hand, $\sigma_{\pm,
\rm ann}$ is much larger and $\sigma_{ee}$ is smaller for the
two-temperature ADAF having a lower $T_e$, 
leading to the low saturation value of $(n_+/n_e)_{\rm eq}
\lesssim 10^{-5}$ as found by Esin et al. 

Therefore, the higher electron temperature of the RIAF model
of Sgr A$^*$ than the ADAF model, which is inferred from the
fitting to observations, is essential to achieve the high
positron production rate to explain the 511 keV emission.

\section{Positron Ejection from Sgr A$^*$}
\label{section:ejection}

\subsection{Annihilation at the Production Site}
Some of the positrons are lost by annihilation around the production
site, producing direct annihilation gamma-ray emission from Sgr
A$^*$. Its spectrum depends on the positron spectrum and gravitational
redshift, and it is thermal with a temperature $T_e \sim 10$ MeV for the
electron-electron scattering.  However, the small value of $(t_{\rm
acc}/t_{\pm, \rm ann}) \sim \dot n_{\pm, \rm ann}/\dot n_+
\sim 4.2 \times 10^{-2} \gamma_+^{-1}$ indicates
that most of the produced positrons are conveyed into the SMBH or
ejected by the wind.  The direct annihilation gamma-ray luminosity from
the pair production site, $L_{\pm, \rm ann} \sim (4\pi r_*^3/3) \ \dot
n_{\pm, \rm ann}$, is then much smaller than the positron production and
ejection rate by the wind, $(n_+/n_e) \dot N_w \sim 4 \pi s r_*^3 \dot
n_+$.

\subsection{Annihilation in the Wind}
Here we check that the positrons trapped in the outflow are not lost by
annihilation before escaping the SMBH gravity. We assume a
constant wind speed
as $\upsilon_{w} \sim \upsilon_{\rm esc}(r_*)$ and the density of the
outflow is $n_{w} \sim \dot N_w / (4 \pi r^2 \upsilon_w)$ for a steady
wind. Then the fraction of positrons that are lost by annihilation
during wind propagation is:
\begin{eqnarray}
\eta_{\rm ann} &=& \int_{r_*}^\infty n_{w} 
\sigma_{\pm, \rm ann} \upsilon_{\pm} \frac{dr}{\upsilon_w} \ ,
\end{eqnarray}
where $\upsilon_{\pm}$ is the relative velocity of electrons and
positrons. Initially both electrons and positrons are relativistic, and
hence $\upsilon_{\pm} \sim c$ and $\sigma_{\pm, \rm ann} \propto
\gamma_\pm^{-1} \sim (\gamma_e \gamma_+)^{-1} \propto r^{4/3}$, since
$\gamma_e$ and $\gamma_+$ scale as $\propto n_w^{1/3} \propto r^{-2/3}$
by adiabatic expansion. Either electrons or positrons become
non-relativistic at $r_{\rm nr} = r_* \min(\gamma_{e*},
\gamma_{+*})^{3/2}$, and the scaling changes as $\sigma_{\pm, \rm ann}
\propto r^{2/3}$ at $r > r_{\rm nr}$, 
where $\gamma_{e*}$ and $\gamma_{+*}$ are the Lorentz
factor at the wind creation site ($r \sim r_*$).  Then, $\sigma_{\pm,
\rm ann} \upsilon_\pm$ becomes constant after both electrons and
positrons become non-relativistic, since $\sigma_{\pm, \rm ann} = \pi
r_e^2 / (\upsilon_\pm/c)$ in the non-relativistic limit (Svensson 1982).
Therefore, the main contribution to the integration comes from $r \sim
r_{\rm nr}$, and we obtain:
\begin{eqnarray}
\eta_{\rm ann} &\sim& 6 n_w(r_{\rm nr}) \sigma_{\pm, \rm ann} (r_{\rm nr}) 
\frac{c}{\upsilon_{\rm esc}(r_*)} \ r_{\rm nr} \\
&\sim& 6 n_w(r_*) \sigma_{\pm, \rm ann}(r_*) 
\frac{c}{\upsilon_{\rm esc}(r_*)} \ r_* \left(\frac{r_{\rm nr}}
{r_*}\right)^{1/3}\\
&=& 1.5 \times 10^{-3} f_{3.5} \gamma_{+*}^{-1} r_3^{1.27} \
\left[\min(\gamma_{e*}, \gamma_{+*})\right]^{1/2} \ .
\end{eqnarray}
This is sufficiently small for any value of $\gamma_{+*}$, and hence we
expect that almost all the positrons produced around the SMBH will
escape from the SMBH gravity field once they are trapped in the
outflowing material.

\subsection{On the Constraint from the Gamma-Ray Background Radiation}
\label{section:background-radiation}

The above results indicate that the annihilation luminosity directly
from Sgr A$^*$ is expected to be much smaller than the bulge 511 keV
line emission, in the phase of the past higher activity.  The
direct annihilation luminosity at present is even much smaller by the
boost factor $f_b$, far below the detection limit of gamma-ray
telescopes in the foreseeable future.

This is important concerning the constraint from the diffuse gamma-ray
background, whose flux is $E \ (dF/dE) \sim 10^{-4} \rm \ photons \
cm^{-2} s^{-1}$ at $\sim$1 MeV within 5$^\circ$ from the Galactic center
(Beacom \& Y\"uksel 2005). If any sources of the positrons
directly emit annihilation gamma-rays before injection of positrons
into ISM, their spectrum would be broad around $\sim$ MeV reflecting the
positron spectrum and gravitational redshift.  The flux from this
process should not violate the observed MeV background, leading to an
upper bound on the annihilation rate as $\lesssim 3.9 \times 10^{41} \
\rm s^{-1}$, assuming no positronium formation for annihilation within
the source. Comparing to the positron annihilation rate of $\sim 1.5
\times 10^{43} \rm s^{-1}$ in ISM, the annihilation rate within the
sources must be $\lesssim$ 2.6\% of the annihilation rate in ISM,
if the sources are concentrated within 5$^\circ$ of the Galactic center.
If the sources are distributed like the observed 511 keV photons,
the constraint becomes $\lesssim$ 10\%, taking into account that
$\sim$24\% of all the 511 keV photons come from the region within
5$^\circ$. 

The model presented here well satisfies this constraint, because of the
current low accretion rate. However, it puts a stringent constraint on
another explanation for the 511 keV emission by accreting black holes,
i.e., LMXBs or microquasars. Theoretical models of pair production in
these objects predict that most ($\sim$90\%) of the produced pairs
annihilate near the production site before injection into ISM (Guessoum
et al. 2006), which is in serious conflict with the above constraint.

\section{Positron Propagation in Interstellar Medium}
\label{section:propagation}

\subsection{Dynamics and Energetics of the Wind Injected into ISM}
\label{section:wind-dynamics}

When positrons escape from the gravitational potential well of the SMBH,
the kinetic energy of the outflow is expected to be dominant compared
with the thermal energy because of the adiabatic cooling, and hence the
positron energy is determined by the bulk Lorentz factor $\Gamma \sim 1$
of the wind, i.e., $\sim$ 1 MeV, as argued in \S
\ref{section:RIAF}. This is important, since too relativistic outflow
would be inconsistent with the observational upper bound on the
injection energy, $\lesssim$ 3 MeV (Beacom \& Y\"uksel 2005; Sizun et
al. 2006).  This bound has been derived by a similar argument in the
previous section (\S \ref{section:background-radiation}), i.e., by
requiring that gamma-rays from in-flight annihilation of positrons
before thermalization in ISM do not make an excess of the diffuse
gamma-ray background radiation.

The wind will sweep up ISM and heat it up by shocks. The ram pressure of
the wind will be balanced with ISM at $r_{p}$, which is determined as
$\dot N_w m_p \upsilon_w / (4 \pi r_p^2) \sim P_{\rm ISM}$, assuming a
quasi-isotropic wind.  Assuming $P_{\rm ISM} = B^2/(8\pi)$ by
interstellar magnetic field of $B \sim 10 \mu$G in the bulge region
(LaRosa et al. 2005; Jean et al. 2006), we find $r_p \sim 3.4 \times
10^2 f_{3.5}^{1/2} r_3^{-0.12} B_{10}^{-1}$ pc, where $B_{10} \equiv
B/(10\mu{\rm G})$.  This radius is comparable with the size of GCL or
EMR, and also with the FWHM of the spatial extent of the 511 keV
emission. Beyond $r \sim r_p$, the shock-heated gas will expand by
thermal pressure. Therefore, even if the wind originally has some
anisotropy, it will not directly appear on a scale larger than $\sim
r_p$.

The observed expansion velocity of $\sim$100 km/s at the GCL/EMR region
indicates an escape time of $t_{\rm esc} \sim 10^6$ yrs from this
region. Then, the wind kinetic energy stored in the GCL region is
\begin{eqnarray}
E_{\rm GCL} &\sim& L_{\rm kin} t_{\rm esc} \\ &\sim& 1.5 \times 10^{55}
f_{3.5} \ r_3^{-0.73} \left(\frac{t_{\rm esc}} {10^6 \ {\rm yrs}}\right)
\ \rm erg \ .
\end{eqnarray}
This is interestingly similar to the energy of the hot gas ($kT \sim 8$
keV) in the Galactic center observed by X-rays (Koyama et al. 1989;
Yamauchi et al. 1990; Muno et al. 2004), $E_{\rm hot} \sim 2.6 \times
10^{54} (r_p / \rm 300 \ pc)^{5/2}$ erg, where I obtained this value from
surface energy density estimated by Muno et al. (2004) assuming the size
and depth to be $\sim r_p$. In fact, the observed size of the hot gas
($\sim 1.8^\circ$ FWHM, Koyama et al. 1989; Yamauchi et al.  1990) is in
good agreement with $r_p$.  Therefore, the large amount of energy stored
in the hard X-ray emitting gas can be explained by the wind
activity. The typical cooling time of the hot gas is $\sim 10^8$ yr
(Muno et al. 2004), which is much longer than $t_{\rm esc}$,
and hence the expansion is adiabatic as argued
by Sofue (2000) and Bland-Hawthorn \& Cohen (2003).  The expansion
velocity ($\sim 100$ km/s) is much lower than the sound velocity of the
hot gas ($\sim 10^3$ km/s), but it is possible if the associated cold
material works as a ballast, as inferred from the infrared emission from
expanding dust (\S \ref{section:outflow}).

\subsection{Positron Propagation Distance}

At $r \gtrsim r_p$, positrons are expected to interact with ISM and
produce the 511 keV emission.  In the hot phase of ISM, which is the
dominant component in the volume filling factor in the bulge region
(Jean et al. 2006), positrons are thermalized in a time scale of $\sim 3
\times 10^6$ yrs and then annihilate in a time scale of $\sim 10^7$ yrs
(Guessoum et al. 2005; Jean et al. 2006). Therefore, the time scale of
the past higher activity inferred from the outflow evidence is
sufficient to explain the 511 keV line emission.  Positrons can
maximally reach $\sim 1$ kpc by the large scale outflow ($\sim$ 100
km/s) within this time scale, being consistent with the observed maximum
extent of the 511 keV emission ($\sim 20^\circ$). Most positrons must
annihilate before traveling this distance, since the spectral analysis
of the 511 keV line indicates that almost all positrons are annihilating
in warm neutral or warm ionized phase of ISM, where the annihilation
time scale is much shorter. Exact spatial profile of the 511 keV line
emission would be determined by the probability of positrons entering
into the warm phase of ISM at $r \gtrsim r_p$.

Diffusion in random magnetic field should also have significant effect.
Jean et al. (2006) estimated the propagation length by quasilinear
diffusion for MeV positrons in the bulge as $\sim 260$ pc for a time
scale of $\sim 3 \times 10^6$ yrs, using a diffusion coefficient $D \sim
9.8 \times 10^{-4} \ \rm kpc^2 Myr^{-1}$ that was derived from $B = 10 \
\mu$G and the Kolmogorov turbulent spectrum. Though the Kolmogorov
spectrum is not valid for cosmic ray propagation in the Galactic disk
(Maurin et al. 2001), this is not significant because a similar
diffusion coefficient of $4.1 \times 10^{-4} \ \rm kpc^2 Myr^{-1}$ is
obtained by using the parameters derived by Maurin et al. (2001) to
explain the cosmic ray data around the Solar system.

A more detailed, quantitative prediction about the 511 keV line
morphology is beyond the scope of this paper, but these considerations
indicate that the model presented here is broadly consistent with the
observed morphology and spatial extent of the 511 keV line emission.

\section{Discussion}
\label{section:discussion}

\subsection{Comparison with Other Models of the Positron Production
from Sgr A$^*$}
\label{section:comparison}

Titarchuk \& Chardonnet (2006) proposed a scenario in which positrons
are produced by annihilation of hard X-ray and $\sim$ 10 MeV photons
around Sgr A$^*$. The hard X-ray photons are emitted from the accretion
activity of the SMBH, while the $\sim 10$ MeV photons are produced by
accretion onto hypothetical small mass black holes (SmMBHs) with a mass
$\sim 10^{17}$ g ($\sim 10^{-16} M_\odot$), which are assumed to be
distributed within $r \sim 10^{2-3} r_s$ of the SMBH. If they are
accreting with the Eddington accretion rate and their density is
comparable with the dark matter, the SmMBHs can supply enough 10 MeV
photons for the required pair production. Such SmMBHs have not yet been
excluded as a candidate of the dark matter, but there is few theoretical
support in contrast to the well-motivated dark matter candidates such as
neutralinos (e.g., Bertone et al. 2004). Furthermore, it is extremely
difficult to supply the accreting material onto such SmMBHs at the
Eddington rate by Bondi accretion; the Bondi accretion rate for a
$10^{17}$ g SmMBH with typical parameters are:
\begin{eqnarray}
\dot M_B \sim 7.9 \times 10^{-23} \ 
\left(\frac{n_e}{\rm cm^{-3}}\right) 
\left(\frac{kT}{1 \ \rm eV}\right)^{-3/2} \ \rm g/s ,
\end{eqnarray}
which should be compared with the Eddington accretion rate of $ \dot
M_{\rm Edd} = 72$ g/s. The difference is almost 24 orders of magnitude,
and it seems quite unlikely that such SmMBHs have the Eddington
accretion rate at any realistic astrophysical circumstances.

Cheng, Chernyshov, \& Dogiel (2006) considered cosmic-ray production by
jet or outflow from the SMBH, which is ejected when stars are captured
by the SMBH. In their scenario, the cosmic rays produce pions by
collisions with ISM, and then positrons are produced by pion decays.
The most important difference of this model from that proposed in this
paper is the high injection energy into ISM ($\gtrsim 30$ MeV) of
pion-decay positrons.  Such a high injection energy is inconsistent with
the observational upper bound on the injection energy, $\lesssim 3$ MeV,
as mentioned in \S \ref{section:wind-dynamics}.

Another problem about positron production from pion decays is the
observed large bulge-to-disk ratio of the 511 keV line emission. We know
that the Galactic gamma-ray background in the GeV band is mainly
composed of pion-decay gamma-rays produced by cosmic-ray interactions in
ISM (Strong et al. 2000, 2004), i.e., the same process with the model of
Cheng et al. The GeV background is clearly associated along with the
Galactic disk, and we have a difficulty to explain why we do not see the
strong disk component of the 511 keV emission if it is produced by the
cosmic-ray interactions.

\subsection{Predictions and Possible Tests by Future Observations}

Although a quantitative prediction of the morphology of the 511 keV
emission is beyond the scope of this paper, it could be more spherically
asymmetric compared with other explanations such as SNe Ia or MeV-mass
dark matter. Asymmetry is expected by the wind anisotropy in the region
of $r \lesssim r_p \sim $ a few degree, or by matter distribution in the
Galaxy at $r \gtrsim r_p$. Asymmetry has not yet been detected in the
observed 511 keV morphology (Kn\"odlseder et al. 2005), but it does not
reject the model presented here because of the limited angular
resolution of the SPI spectrometer of INTEGRAL ($\sim 3^\circ$ FWHM) and
the large uncertainty of theoretical expectation for the
anisotropy. Future observational studies on the 511 keV morphology with
better angular resolution might, however, detect larger asymmetry than
that expected for the other explanations.

Another prediction of the proposed scenario is that it is extremely
difficult to detect 511 keV emission in regions other than the Galactic
bulge from the source population that is responsible for the bulge
component. SMBHs are generally found in galaxies having bulges, and the
nearest SMBH is probably M31$^*$ in the Andromeda galaxy.  The X-ray
luminosity of M31$^*$ is $\sim 10^{36}$ erg/s (Garcia et al. 2005),
which is about $10^3$ times larger than the quiescent phase of Sgr
A$^*$, but $10^3$ times smaller than the past higher activity phase. If
this X-ray luminosity reflects the typical activity averaged over a
time scale of $\sim 10^7$ yr, we expect that the 511 keV luminosity of
the M31 bulge is much fainter than that of the Galaxy. The 511 keV
emission has been detected at $\sim 50 \sigma$ level (Kn\"odlseder et
al. 2005), and considering the distance to M31 (770 kpc), a large
improvement of the sensitivity is required.

On the other hand, there is a better chance to detect 511 keV emission
from regions other than the Galactic bulge, for some other models of the
bulge 511 keV emission (e.g., Kn\"odlseder et al. 2005). For example, if
the origin is SNe Ia, improved instruments in the near future will
detect 511 keV line emission from nearby supernova remnants. 
In fact, an interesting limit on the positron escape fraction
from SN 1006 has already been obtained (Kalemci et al. 2006). If the
origin is the MeV-mass dark matter annihilation, we expect 511 keV line
emission from nearby dwarf galaxies by a modest improvement of the
sensitivity. If future negative results ruled out the other
explanations of the 511 keV emission, it would strengthen the case for
the scenario presented here.

It should also be noted that the standard prediction for positron
production from SNe Ia is only one order of magnitude short of that
required to explain the bulge 511 keV emission (Prantzos
2006). Therefore, just a detection of 511 keV emission from a nearby SN
Ia does not confirm SNe Ia as the origin of the bulge component, but a
close examination of the positron production efficiency would be
required.

\section{Conclusion}

In this paper, I have shown that the several independent lines of
evidence for the past higher activity of the Galactic center (i.e.,
higher X-ray luminosity and large scale outflows) can quantitatively be
explained by the RIAF model of Sgr A$^*$, in which energetic outflow
plays an essential role.  A single increased accretion rate from the
current value explains both the past high X-ray luminosity and kinetic
luminosity of outflow inferred from observations. The required boost
factor of the accretion rate is about $10^{3-4}$ for a time scale of
$\sim 10^7$ yrs in the past. I have shown that this accretion rate and
its duration are naturally expected in the environment of the Galactic
center.  The outflow is energetic enough to supply the heat to keep the
hot $\sim$ 8 keV plasma observed in the Galactic center region, for
which there have been no good explanation.  The current low accretion
rate is likely caused by a sudden destruction of the accretion flow when
the shell of the supernova remnant Sgr A East passed through the SMBH
about 300 yrs ago. The chance probability of observing Sgr A$^*$ in such
a low activity phase is estimated to be $\sim 0.5$\%.

I then estimated the production rate of positrons during the high
activity phase, which are created in the region around the event horizon
and then ejected by the outflow. The rate is found to be comparable with
that required to explain the 511 keV line emission toward the Galactic
bulge.  I considered three processes of $e^\pm$ pair-production via
electron-electron scattering, photon-electron collision, and
photon-photon annihilation, and interestingly all the three processes
give a similar positron production rate. 

Therefore the model presented here gives a new and natural explanation
for the 511 keV line emission toward the Galactic bulge. The favorable
aspects of this model are: (1) the correct positron production rate, (2)
the large bulge-to-disk ratio and correct spatial extent in the bulge,
(3) (i) negligible annihilation near the positron production site before
injection into ISM, and (ii) the injection energy of $\sim$ MeV, both of
which are important to meet the constraint from the MeV gamma-ray
background, and (4) no exotic assumptions or parameters.  To my
knowledge, the other explanations for the 511 keV emission proposed
so far do not satisfy all of these.

Anisotropy of the morphology of the 511 keV emission larger than that
expected for other models would be a signature for the model proposed
here, which might be revealed by future observations with better angular
resolutions. Detection of 511 keV lines from centers of nearby galaxies
by the same mechanism will be difficult even in the foreseeable
future. In contrast, some other scenarios predict detectable 511 keV
lines in directions other than the Galactic center by a modest
improvement of the sensitivity.

I would like to thank an anonymous referee for useful comments.
This work was supported by the Grant-in-Aid for the 21st Century COE
``Center for Diversity and Universality in Physics'' from the Ministry
of Education, Culture, Sports, Science and Technology (MEXT) of Japan.

\onecolumn

\begin{table}
  \caption{Summary of observations versus the model} 
  \label{table:summary}
\begin{center}
\begin{tabular}{lllll}
  \hline\hline
  Observation & Scale (pc) & Requirement$^*$ 
      & Prediction$^\dagger$ & Section \\
  \hline
  \multicolumn{5}{c}{Evidence for higher past X-ray luminosity} \\
  \hline
  X-ray reflection nebulae & 100 & $L_X \sim 3 \times 10^{39}$ erg/s 
     & $5 \times 10^{39} f_{3.5}$  & \ref{section:past-X-ray-luminosity} \\
  Ionized halo & 10 & $L_X \sim 10^{40}$ erg/s 
     & $5 \times 10^{39} f_{3.5}$   &
     \ref{section:past-X-ray-luminosity} \\
  \hline
  \multicolumn{5}{c}{Evidence for past outflow activity} \\
  \hline
  Galactic center lobe (GCL) & 200 & $L_{\rm kin} \sim 3 \times 
   10^{41}$ erg/s  & $4.7 \times 10^{41} f_{3.5}$   &
      \ref{section:outflow} \\
  Expanding molecular ring (EMR) & 300 & $L_{\rm kin} \sim 1 \times 10^{42}$
   erg/s  &$4.7 \times 10^{41} f_{3.5}$   & \ref{section:outflow} \\
  North polar spur (NPS) & $4 \times 10^3$ & $L_{\rm kin} \sim 1 \times 
   10^{41}$ erg/s  & $4.7 \times 10^{41} f_{3.5}$ &
   \ref{section:outflow} \\
  8 keV diffuse gas  & 300  & $E_{\rm hot} \sim 2.6 \times 10^{54}$ erg 
     & $1.5 \times 10^{55} f_{3.5}$   & \ref{section:outflow},
     \ref{section:wind-dynamics}  \\
  \hline
  \multicolumn{5}{c}{Evidence for pair production} \\
  \hline
  Bulge 511 keV line emission  & 600  & $\dot N_+ \sim 1.5 \times 10^{43} 
    \ \rm s^{-1}$  
     &   & \\
   & & (production by $ee$) & $3.7 \times 10^{42} f_{3.5}^2$ & 
     \ref{section:electron-electron} \\
   & & (production by $e\gamma$) & $8.1 \times 10^{41} 
      f_{3.5}^3$ & \ref{section:electron-photon}  \\
   & & (production by $\gamma \gamma$) & $1.1 \times 10^{42} 
      f_{3.5}^4$ & \ref{section:photon-photon} \\
  \hline \hline
\end{tabular}
\end{center}
$^*$X-ray luminosity ($L_X$), kinetic luminosity
  $L_{\rm kin}$, energy stored in the hot gas ($E_{\rm hot}$),
  or positron production rate ($\dot N_+$) required
  to explain the observations. \\
$^\dagger$The dependence on parameters other than the accretion rate 
 is omitted, where $f_{3.5} = f_b/10^{3.5}$ and
 $f_b = \dot M_{\rm past} / \dot M_{\rm present}$. 
\end{table}

\begin{thebibliography}{}

\bibitem[Aharonian et al.(2004)]{2004A&A...425L..13A} Aharonian, F., et 
al.\ 2004, \aap, 425, L13 

\bibitem[Aharonian \& Neronov(2005)]{2005ApJ...619..306A} Aharonian, F.,
\& Neronov, A.\ 2005, \apj, 619, 306 

\bibitem[Anantharamaiah et al.(1999)]{1999ASPC..186..422A}
Anantharamaiah, K.~R., Pedlar, A., \& Goss, W.~M.\ 1999, ASP
Conf.~Ser.~186: The Central Parsecs of the Galaxy, 186, 422

\bibitem[Baganoff et al.(2003)]{2003ApJ...591..891B} Baganoff, F.~K., et 
al.\ 2003, \apj, 591, 891 

\bibitem[Beacom \& Yuksel(2005)]{2005astro.ph.12411B} Beacom, J.~F., \& 
Yuksel, H.\ 2005, ArXiv Astrophysics e-prints, arXiv:astro-ph/0512411 

\bibitem[Bertone et al.(2004)]{2004PhR...405..279B} Bertone, G., Hooper, 
D., \& Silk, J.\ 2004, \physrep, 405, 279 

\bibitem[Bjoernsson et al.(1996)]{1996ApJ...467...99B} Bj\"ornsson, G., 
Abramowicz, M.~A., Chen, X., \& Lasota, J.-P.\ 1996, \apj, 467, 99 

\bibitem[Blandford \& Begelman(1999)]{1999MNRAS.303L...1B} Blandford, 
R.~D., \& Begelman, M.~C.\ 1999, \mnras, 303, L1 

\bibitem[Bland-Hawthorn \& Cohen(2003)]{2003ApJ...582..246B} 
Bland-Hawthorn, J., \& Cohen, M.\ 2003, \apj, 582, 246 

\bibitem[Cheng et al. (2006)]{astro-ph/0603659} 
Cheng, K.S., Chernyshov, D.O., \& Dogiel, V.A.
2006, to appear in ApJ, astro-ph/0603659

\bibitem[Cuadra et al.(2006)]{2006MNRAS.366..358C} Cuadra, J.,
Nayakshin, S., Springel, V., \& di Matteo, T.\ 2006, \mnras, 366, 358 

\bibitem[Diehl et al.(2006)]{2006Natur.439...45D} Diehl, R., et al.\
2006, \nat, 439, 45

\bibitem[Eisenhauer et al.(2003)]{2003ApJ...597L.121E} Eisenhauer, F., 
Sch{\"o}del, R., Genzel, R., Ott, T., Tecza, M., Abuter, R., Eckart, A., \& 
Alexander, T.\ 2003, \apjl, 597, L121 

\bibitem[Esin et al.(1997)]{1997ApJ...489..865E} Esin, A.~A., McClintock, 
J.~E., \& Narayan, R.\ 1997, \apj, 489, 865 

\bibitem[Falcke \& Biermann(1999)]{1999A&A...342...49F} Falcke, H., \& 
Biermann, P.~L.\ 1999, \aap, 342, 49 

\bibitem[Gallo et al.(2003)]{2003MNRAS.344...60G} Gallo, E., Fender, R.~P., 
\& Pooley, G.~G.\ 2003, \mnras, 344, 60 

\bibitem[Gallo et al.(2005)]{2005Natur.436..819G} Gallo, E., Fender, R., 
Kaiser, C., Russell, D., Morganti, R., Oosterloo, T., \& Heinz, S.\ 2005, 
\nat, 436, 819 

\bibitem[Garcia et al.(2005)]{2005ApJ...632.1042G} Garcia, M.~R., Williams, 
B.~F., Yuan, F., Kong, A.~K.~H., Primini, F.~A., Barmby, P., Kaaret, P., \& 
Murray, S.~S.\ 2005, \apj, 632, 1042 

\bibitem[Genzel et al.(1997)]{1997MNRAS.291..219G} Genzel, R., Eckart,
A., Ott, T., \& Eisenhauer, F.\ 1997, \mnras, 291, 219

\bibitem[Guessoum et al.(2005)]{2005A&A...436..171G} Guessoum, N., Jean, 
P., \& Gillard, W.\ 2005, \aap, 436, 171 

\bibitem[Guessoum et al.(2006)]{2006astro.ph..7296G} Guessoum, N., Jean, 
P., \& Prantzos, N.\ 2006, to appear in A\&A (astro-ph/0607296)

\bibitem[Ho(2004)]{2004cbhg.symp..292H} Ho, L.~C.~W.\ 2004, Coevolution
of Black Holes and Galaxies, 292 (astro-ph/0401527)

\bibitem[Hooper \& Dingus(2004)]{2004PhRvD..70k3007H} Hooper, D., \& 
Dingus, B.~L.\ 2004, \prd, 70, 113007

\bibitem[Jean et al.(2006)]{2006A&A...445..579J} Jean, P.,
Kn{\"o}dlseder, J., Gillard, W., Guessoum, N., Ferri{\`e}re, K.,
Marcowith, A., Lonjou, V., \& Roques, J.~P.\ 2006, \aap, 445, 579

\bibitem[Kaifu et al.(1972)]{1972Natur.238..105K} Kaifu, N., Kato, T., \& 
Iguchi, T.\ 1972, \nat, 238, 105 

\bibitem[Kalemci et al.(2006)]{2006ApJ...640L..55K} Kalemci, E., Boggs, 
S.~E., Milne, P.~A., \& Reynolds, S.~P.\ 2006, \apjl, 640, L55 

\bibitem[Kato et al.(1998)]{1998bhad.conf.....K} Kato, S., Fukue, J., \& 
Mineshige, S.\ 1998, Black-hole accretion disks.~ Edited by Shoji Kato,
Jun Fukue, and Sin Mineshige.~ Publisher: Kyoto, Japan: Kyoto University
Press, 1998


\bibitem[Kn{\"o}dlseder et al.(1999)]{1999A&A...344...68K}
Kn{\"o}dlseder, J., et al.\ 1999, \aap, 344, 68 

\bibitem[Kn{\"o}dlseder et al.(2005)]{2005A&A...441..513K}
Kn{\"o}dlseder, J., et al.\ 2005, \aap, 441, 513 

\bibitem[Koyama et al.(1989)]{1989Natur.339..603K} Koyama, K., Awaki, H., 
Kunieda, H., Takano, S., \& Tawara, Y.\ 1989, \nat, 339, 603 

\bibitem[Koyama et al.(1996)]{1996PASJ...48..249K} Koyama, K., Maeda,
Y., Sonobe, T., Takeshima, T., Tanaka, Y., \& Yamauchi, S.\ 1996, \pasj,
48, 249

\bibitem[Koyama et al.(2006)]{2006astro.ph..9215K} Koyama, K., et al.\ 
2006, to appear in PASJ, arXiv:astro-ph/0609215 

\bibitem[Kusunose \& Mineshige(1996)]{1996ApJ...468..330K} Kusunose, M., \& 
Mineshige, S.\ 1996, \apj, 468, 330 

\bibitem[LaRosa et al.(2005)]{2005ApJ...626L..23L} LaRosa, T.~N., Brogan, 
C.~L., Shore, S.~N., Lazio, T.~J., Kassim, N.~E., \& Nord, M.~E.\ 2005, 
\apjl, 626, L23 

\bibitem[Le \& Becker(2004)]{2004ApJ...617L..25L} Le, T., \& Becker,
P.~A.\ 2004, \apjl, 617, L25 

\bibitem[Maeda et al.(2002)]{2002ApJ...570..671M} Maeda, Y., et al.\
2002, \apj, 570, 671 

\bibitem[Martini(2004)]{2004cbhg.symp..169M} Martini, P.\ 2004, Coevolution 
of Black Holes and Galaxies, ed. L. C. Ho (Cambridge: Cambridge
University Press) 169 

\bibitem[Maurin et al.(2001)]{2001ApJ...555..585M} Maurin, D., Donato,
F., Taillet, R., \& Salati, P.\ 2001, \apj, 555, 585 

\bibitem[Mayer-Hasselwander et al.(1998)]{1998A&A...335..161M} 
Mayer-Hasselwander, H.~A., et al.\ 1998, \aap, 335, 161 

\bibitem[Mineshige et al.(1995)]{1995ApJ...445L..43M} Mineshige, S., 
Kusnose, M., \& Matsumoto, R.\ 1995, \apjl, 445, L43 

\bibitem[Muno et al.(2004)]{2004ApJ...613..326M} Muno, M.~P., et al.\
2004, \apj, 613, 326 

\bibitem[Murakami et al.(2000)]{2000ApJ...534..283M} Murakami, H.,
Koyama, K., Sakano, M., Tsujimoto, M., \& Maeda, Y.\ 2000, \apj, 534, 283 

\bibitem[Murakami et al.(2001a)]{2001ApJ...558..687M} Murakami, H., Koyama, 
K., \& Maeda, Y.\ 2001a, \apj, 558, 687 

\bibitem[Murakami et al.(2001b)]{2001ApJ...550..297M} Murakami, H., Koyama, 
K., Tsujimoto, M., Maeda, Y., \& Sakano, M.\ 2001b, \apj, 550, 297 

\bibitem[Narayan et al.(1998)]{1998tbha.conf..148N} Narayan, R.,
Mahadevan, R., \& Quataert, E.\ 1998, Theory of Black Hole Accretion
Disks, 148 (astro-ph/9803141)

\bibitem[Narayan \& McClintock(2005)]{2005ApJ...623.1017N} Narayan, R., \& 
McClintock, J.~E.\ 2005, \apj, 623, 1017 

\bibitem[Narayan \& Quataert(2005)]{2005Sci...307...77N} Narayan, R., \& 
Quataert, E.\ 2005, Science, 307, 77 

\bibitem[Ohsuga et al.(2005)]{2005ApJ...627..782O} Ohsuga, K., Kato, Y., \& 
Mineshige, S.\ 2005, \apj, 627, 782 

\bibitem[Paumard et al.(2006)]{2006ApJ...643.1011P} Paumard, T., et al.\ 
2006, \apj, 643, 1011 

\bibitem[Pedlar et al.(1989)]{1989ApJ...342..769P} Pedlar, A., 
Anantharamaiah, K.~R., Ekers, R.~D., Goss, W.~M., van Gorkom, J.~H., 
Schwarz, U.~J., \& Zhao, J.-H.\ 1989, \apj, 342, 769 

\bibitem[Prantzos \& Diehl(1996)]{1996PhR...267....1P} Prantzos, N., \& 
Diehl, R.\ 1996, \physrep, 267, 1 

\bibitem[Prantzos(2006)]{2006A&A...449..869P} Prantzos, N.\ 2006, \aap, 
449, 869 

\bibitem[Revnivtsev et al.(2004)]{2004A&A...425L..49R} Revnivtsev,
M.~G., et al.\ 2004, \aap, 425, L49 

\bibitem[Revnivtsev et al.(2006)]{2006A&A...452..169R} Revnivtsev, M., 
Sazonov, S., Gilfanov, M., Churazov, E., \& Sunyaev, R.\ 2006, \aap,
452, 169 

\bibitem[Salamon \& Stecker(1998)]{1998ApJ...493..547S} Salamon, M.~H., \& 
Stecker, F.~W.\ 1998, \apj, 493, 547 

\bibitem[Sch{\"o}del et al.(2003)]{2003ApJ...596.1015S} Sch{\"o}del, R., 
Ott, T., Genzel, R., Eckart, A., Mouawad, N., \& Alexander, T.\ 2003,
\apj, 596, 1015 


\bibitem[Scoville(1972)]{1972ApJ...175L.127S} Scoville, N.~Z.\ 1972, \apjl, 
175, L127 

\bibitem[Sizun et al.]{} Sizun, P., Cass\'e, M., \& Schanne, S.
 2006, submitted to Phys. Rev. D. (astro-ph/0607374)

\bibitem[Sofue(2000)]{2000ApJ...540..224S} Sofue, Y.\ 2000, \apj, 540,
224 

\bibitem[Strong et al.(2000)]{2000ApJ...537..763S} Strong, A.~W., 
Moskalenko, I.~V., \& Reimer, O.\ 2000, \apj, 537, 763 

\bibitem[Strong et al.(2004)]{2004ApJ...613..962S} Strong, A.~W., 
Moskalenko, I.~V., \& Reimer, O.\ 2004, \apj, 613, 962 

\bibitem[Svensson(1982)]{1982ApJ...258..335S} Svensson, R.\ 1982, \apj, 
258, 335 

\bibitem[Svensson(1984)]{1984MNRAS.209..175S} Svensson, R.\ 1984, \mnras, 
209, 175 

\bibitem[Titarchuk \& Chardonnet(2006)]{2006ApJ...641..293T} Titarchuk,
L., \& Chardonnet, P.\ 2006, \apj, 641, 293 

\bibitem[Yamauchi et al.(1990)]{1990ApJ...365..532Y} Yamauchi, S., Kawada, 
M., Koyama, K., Kunieda, H., \& Tawara, Y.\ 1990, \apj, 365, 532 

\bibitem[Yuan et al.(2002)]{2002A&A...383..854Y} Yuan, F., Markoff, S.,
\& Falcke, H.\ 2002, \aap, 383, 854 

\bibitem[Yuan et al.(2003)]{2003ApJ...598..301Y} Yuan, F., Quataert, E.,
\& Narayan, R.\ 2003, \apj, 598, 301 (YQN03)

\bibitem[Yuan et al.(2004)]{2004ApJ...606..894Y} Yuan, F., Quataert, E.,
\& Narayan, R.\ 2004, \apj, 606, 894  (YQN04)

\bibitem[Yusef-Zadeh et al.(2000)]{2000Sci...287...85Y} Yusef-Zadeh, F., 
Melia, F., \& Wardle, M.\ 2000, Science, 287, 85 

\bibitem[Zdziarski(1982)]{1982A&A...110L...7Z} Zdziarski, A.~A.\ 1982, 
\aap, 110, L7 

\bibitem[Zdziarski(1985)]{1985ApJ...289..514Z} Zdziarski, A.~A.\ 1985, 
\apj, 289, 514 

\end{thebibliography}
\end{document}